\documentclass[conference]{IEEEtran}
\usepackage{cite}
\usepackage{amsmath,amssymb,amsfonts}
\usepackage{algorithmic}
\usepackage{graphicx}
\graphicspath{{./figures/}}
\usepackage[caption=false,font=footnotesize]{subfig}
\usepackage{textcomp}
\usepackage{xcolor}
\usepackage{url}
\usepackage{multirow}
\usepackage{booktabs}
\usepackage{array}
\usepackage{tikz}
\usetikzlibrary{arrows.meta,positioning}
\def\BibTeX{{\rm B\kern-.05em{\sc i\kern-.025em b}\kern-.08em
    T\kern-.1667em\lower.7ex\hbox{E}\kern-.125emX}}
\begin{document}

\title{A Training-Efficient Transformer-Based Anti-Spoofing Network for Logical Access in ASVspoof~5}

\author{
  \IEEEauthorblockN{
    Sidan Yin\textsuperscript{1},
    Bo Zhao\textsuperscript{2},
  }
  \\
  \IEEEauthorblockA{\textsuperscript{1}San Domenico School, San Anselmo, CA, United States}
  \IEEEauthorblockA{\textsuperscript{2}University of Washington, Seattle, WA, United States}
}

\maketitle

\begin{abstract}
Synthetic and manipulated speech can reduce the reliability of automatic speaker verification systems, so anti-spoofing methods need to be both accurate and efficient in training and inference.
This paper focuses on the ASVspoof~5 Track~1 closed condition, where standard cross-entropy training may not give enough attention to hard trials and is not directly aligned with ranking- and threshold-based evaluation metrics.
We propose TFPARN, a Transformer-based focal--pairwise attentive ranking network.\footnote{Code is available at \url{https://github.com/SomeB1oody/TFPARN}}
The system extracts log-Mel features from speech, uses a Transformer encoder to model frame-level information, applies attention pooling to obtain utterance-level representations, and is trained with a combination of focal classification loss and pairwise ranking loss.
RawBoost augmentation is used during training, and test-time augmentation is applied during evaluation to improve robustness.
Compared with re-implemented AASIST and RawNet2 baselines under the same protocol, TFPARN achieves the best results, with a minDCF of 0.2430 and an EER of 12.52\%.
Ablation experiments further show that the pairwise loss, focal loss, and attention pooling all improve performance.
TFPARN also uses the lowest inference memory among the compared systems, at 1.4~GB, runs at about 0.79~ms per utterance, and reaches its best checkpoint in less training time than AASIST.
These results show that TFPARN provides a good balance between detection accuracy and computational cost for logical access anti-spoofing.
\end{abstract}

\begin{IEEEkeywords}
Speech anti-spoofing, speech deepfake detection, ASVspoof~5, Transformer, focal loss, pairwise ranking, training efficiency
\end{IEEEkeywords}

\section{Introduction}
Advances in artificial intelligence (AI) have made generated speech increasingly difficult to distinguish from genuine recordings \cite{refTacotron2}. 
The same technology also makes voice forgery easier and threatens the reliability of automatic speaker verification (ASV) systems.

The ASVspoof challenge series was established to address such threats arising from synthetic and manipulated speech \cite{refASV2015, refASV2017, refASV2019, refASV2021, refASV5}. 
It encourages the development of general-purpose countermeasures, such as spoofing and deepfake detection systems that stay reliable even against attacks from new or unseen spoofing algorithms. 
The latest edition, ASVspoof~5, consists of two tracks: Track~1 focuses on stand-alone spoofing and speech deepfake detection, whereas Track~2 targets spoofing-robust automatic speaker verification (SASV) \cite{refASV5}. 
This work concentrates on Track~1.

ASVspoof~5 Track~1 presents several challenges. 
Spoofing trials vary substantially in difficulty: some utterances contain obvious artifacts and are relatively easy to detect, whereas others are highly natural.
Conventional cross-entropy (CE) training tends to give too little weight to these hard examples once most samples are already well classified. 
The primary evaluation metrics of Track~1, such as EER and minDCF, are ranking-based rather than purely classification-based \cite{refASV5}, and CE does not explicitly enforce a globally favorable score ordering. 
Spoofing traces are also often local and sparse in time and frequency, so simple global aggregation operations may dilute the most informative regions.

Practical deployment imposes computational constraints on ASV systems.
Voice verification devices often have long service lifecycles and limited hardware, but users want verification to complete with low latency.
The system must also be re-trained periodically as new data are collected, so training cost largely determines how quickly it can be updated.
Training and inference efficiency therefore bear directly on the cost of deploying and maintaining an ASV system.

However, existing anti-spoofing systems do not yet offer a model that balances detection performance against computational efficiency. 
Representative models such as AASIST \cite{refAASIST} and RawNet2 \cite{refRawNet2} are not explicitly designed to jointly optimize training cost and inference efficiency. 

To address this issue, this paper proposes the Transformer-based Focal-Pairwise Attentive Ranking Network (TFPARN) for ASVspoof~5 Track~1. 
TFPARN uses a Transformer encoder \cite{refTransformer} to model long-range temporal dependencies in log-Mel sequences, with attentive pooling to emphasize frames that are more likely to contain spoofing clues. 
TFPARN adopts focal loss, combined with pairwise ranking loss, to better align training with ranking-sensitive metrics such as EER and minDCF \cite{refFocal, refPairwise}. 
RawBoost waveform-level augmentation \cite{refRawBoost} and test-time augmentation (TTA) are used to improve robustness.

This work does not aim solely for the best detection score; 
its goal is to strike a better balance between anti-spoofing performance and computational efficiency under the ASVspoof~5 Track~1 closed condition. 
We analyze detection performance, training efficiency, and inference efficiency of TFPARN. 

The main contributions of this work are summarized as follows:
\begin{itemize}
    \item We propose TFPARN, a Transformer-based anti-spoofing architecture for the ASVspoof~5 Track~1 closed condition, designed to balance detection performance against training and inference cost.
    \item We use attentive temporal pooling to focus on local clues of fake speech, focal loss to emphasize difficult trials, and pairwise ranking loss to improve alignment with ranking-oriented anti-spoofing metrics.
    \item We compare TFPARN with re-implemented baseline models under a unified protocol, evaluating the resulting systems by performance metrics and computational efficiency.
    \item We show that TFPARN offers a favorable trade-off between anti-spoofing performance and the efficiency of training and inference.
\end{itemize}

\section{ASVspoof~5 Track~1 Closed Condition Task}

\subsection{Task Description and Closed Condition}

ASVspoof~5 evaluates countermeasure (CM) algorithms that discriminate bonafide speech from spoofed/deepfake speech \cite{refASV5}. 
The challenge has two tracks. 
This paper focuses on Track~1, where the system receives telephony/VoIP speech and outputs a single real-valued detection score, with larger values indicating stronger support for the bonafide class.

ASVspoof~5 defines open and closed conditions, and TFPARN is trained under the closed condition \cite{refASV5}. 
In the closed condition, systems may only use the official ASVspoof~5 training partition, with no external data or pre-trained models. 
We conduct all experiments under the Track~1 closed condition.

\subsection{Dataset and Protocols}

The ASVspoof~5 data are built from the English subset of the Multilingual LibriSpeech (MLS) corpus \cite{refMLS}. 
The organizers provide disjoint train, dev, and eval partitions with non-overlapping speakers and spoofing attacks, which increases the difficulty and tests generalization to unseen speakers and attacks.
Based on the official Track~1 protocol files, the approximate data scale and class ratios are summarized in Table~\ref{tab1}.

\begin{table}[htbp]
\caption{ASVspoof~5 data statistics (Track~1)}
\begin{center}
\begin{tabular}{cccccc}
\toprule
\textbf{Subset} & \textbf{Speakers} & \textbf{Attacks} & \textbf{Utterances} & \textbf{Spoof} & \textbf{Bonafide} \\
\midrule
Train & 400 & 8 & 182,357 & 163,560 & 18,797 \\
Dev & 785 & 8 & 140,950 & 109,616 & 31,334 \\
\bottomrule
\end{tabular}
\label{tab1}
\end{center}
\end{table}

As shown, the Track~1 train and dev subsets are completely disjoint in speakers (400 vs.~785) and cover 8 attack types each. 
The spoofing implementations are also disjoint across train, dev, and eval (8, 8, and 16 attacks, respectively), which further increases task difficulty.

\subsection{Evaluation Metrics}

For Track~1, the official evaluation plan defines the following metrics:
\begin{itemize}
  \item Minimum Detection Cost Function (minDCF, primary metric);
  \item Equal Error Rate (EER);
  \item Cost of Log-Likelihood Ratios (Cllr) \cite{refCllr};
  \item Actual Detection Cost Function (actDCF).
\end{itemize}

All metrics follow the official ASVspoof~5 Evaluation Plan \cite{refASV5}.
For Track~1, minDCF and actDCF use spoof prior \( \pi_{\mathrm{spf}} = 0.05 \) with costs \( C_{\mathrm{miss}} = 1 \) and \( C_{\mathrm{fa}} = 10 \). 
EER and minDCF are ranking/threshold-sensitive metrics under the Track~1 protocol. 
Since cross-entropy does not optimize global score ordering under the target prior and costs, we additionally combine pairwise ranking loss with focal loss (for hard-sample emphasis) to better align training with EER and minDCF \cite{refFocal, refPairwise}.

\section{System Overview}

\subsection{Overall Architecture of TFPARN}

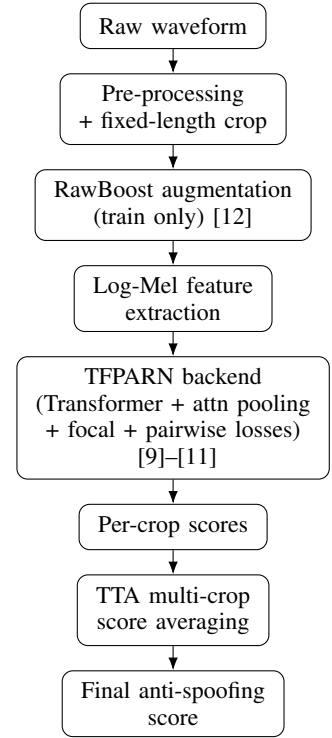
\begin{figure}[t]
  \centering
  \begin{tikzpicture}[
      node distance=3.2mm and 2.0mm,
      box/.style={draw, rounded corners, align=center, font=\small, minimum height=6mm, inner xsep=2.5mm, inner ysep=1.5mm},
      arrow/.style={-Latex, line width=0.4pt}
    ]
    \node[box] (raw) {Raw waveform};
    \node[box, below=of raw] (prep) {Pre-processing\\+ fixed-length crop};
    \node[box, below=of prep] (rb) {RawBoost augmentation\\(train only) \cite{refRawBoost}};
    \node[box, below=of rb] (logmel) {Log-Mel feature\\extraction};
    \node[box, below=of logmel] (tfparn) {TFPARN backend\\(Transformer + attn pooling\\+ focal + pairwise losses)\\\cite{refTransformer, refFocal, refPairwise}};
    \node[box, below=of tfparn] (score) {Per-crop scores};
    \node[box, below=of score] (tta) {TTA multi-crop\\score averaging};
    \node[box, below=of tta] (final) {Final anti-spoofing\\score};

    \draw[arrow] (raw) -- (prep);
    \draw[arrow] (prep) -- (rb);
    \draw[arrow] (rb) -- (logmel);
    \draw[arrow] (logmel) -- (tfparn);
    \draw[arrow] (tfparn) -- (score);
    \draw[arrow] (score) -- (tta);
    \draw[arrow] (tta) -- (final);
  \end{tikzpicture}
  \caption{Overall processing pipeline of TFPARN for ASVspoof~5 Track~1.}
  \label{fig:pipeline_overview}
\end{figure}

The overall processing pipeline of the proposed system is summarized in Fig.~\ref{fig:pipeline_overview}.

During training, each input utterance is first resampled to 16~kHz, converted to a single-channel waveform of fixed duration 4.0~s, and amplitude-normalized. 
With a 50\% probability, RawBoost waveform-level augmentation is then applied \cite{refRawBoost}.
The resulting signal is transformed into a log-Mel spectrogram sequence via a short-time Fourier transform (STFT) followed by a Mel filterbank, and fed into the TFPARN model. 
The model outputs binary logits for the ``spoof/bonafide'' classes. 
The parameters are optimized jointly using a focal classification loss and a pairwise ranking loss, emphasizing confusing samples and hard spoof-bonafide pairs\cite{refFocal, refPairwise}.

On the development and evaluation sets, the system generates multiple fixed-length crops per utterance for TTA.
Each crop is passed through the same TFPARN model to obtain a set of detection scores, which are then averaged to reduce the variance induced by any single crop. 
The averaged score is used as the final score to compute evaluation metrics such as EER and minDCF.

The core of this work is the TFPARN model itself. 
RawBoost augmentation and TTA multi-crop scoring are auxiliary components that improve robustness under the evaluation conditions of ASVspoof~5 Track~1.

\subsection{Front-end Features and Preprocessing}

\subsubsection{Waveform Pre-processing and Cropping Strategy}

All utterances are first resampled to 16~kHz and converted to single-channel floating-point waveforms; for multi-channel recordings, the channel-wise average is taken. 
The waveform amplitude is then normalized so that its samples lie in the range \([-1, 1]\), reducing the impact of differences in recording level.

Each utterance is unified to a fixed duration of 4.0 s, corresponding to  \(T_{\text{target}} = 4.0 \times 16000 = 64000\) samples. 
Given an input waveform of length \(T\), the fixed-length processing strategy is as follows:
\begin{itemize}
  \item \textit{If \(T > T_{\text{target}}\)}: On the training set, a random start index \(t_0 \sim \text{Uniform}(0, T - T_{\text{target}})\) is sampled and a segment of length \(T_{\text{target}}\) is cropped to increase data diversity; 
  Center crop is used on the Dev/Eval sets to ensure evaluation consistency.

  \item \textit{If \(T < T_{\text{target}}\)}: The entire waveform is repeatedly concatenated along the time axis until the length exceeds \(T_{\text{target}}\), yielding a repeated waveform of length \(T'\). 
  On the training set, a random segment of length \(T_{\text{target}}\) is cropped with the start index sampled from \([0, T' - T_{\text{target}}]\); on the Dev/Eval sets, a center crop is taken from the repeated waveform.
\end{itemize}

\subsubsection{RawBoost Waveform-level Augmentation}

To improve robustness against varying spoofing conditions and channel mismatch, RawBoost waveform-level data augmentation is adopted during training.
It is applied only to the training set with probability \(p_{\text{rb}} = 0.5\) \cite{refRawBoost}.
RawBoost consists of the following three algorithms.

\begin{itemize}
    \item \textit{Convolutive noise (Algorithm 1):}
    A randomly generated short FIR filter \(h\) is convolved with
    the waveform:
    \begin{equation}
    \tilde{x} = x \ast h,
    \end{equation}
    optionally followed by
    \begin{equation}
    x' = \tanh(\alpha \tilde{x}),
    \end{equation}
    to emulate channel distortion.

    \item \textit{IIR filtering (Algorithm 2):}
    A low-pass, high-pass, or band-pass Butterworth filter
    is randomly selected. The waveform is filtered as
    \begin{equation}
    x' = \text{IIRFilter}(x; b, a),
    \end{equation}
    where \(b\) and \(a\) are the filter coefficients.

    \item \textit{Stationary additive noise (Algorithm 3):}
    A Gaussian noise sequence \(n\) is generated and scaled
    according to \(\mathrm{SNR} \in [10,40]\) dB:
    \begin{equation}
    \begin{aligned}
    x' &= x + n, \\
    \frac{\mathbb{E}[x^2]}{\mathbb{E}[n^2]}
    &= 10^{\mathrm{SNR}/10}.
    \end{aligned}
    \end{equation}
\end{itemize}

During training, each utterance is augmented by randomly selecting one of the above three algorithms with probability \(p_{\text{rb}}\).
RawBoost is completely disabled on the Dev and Eval sets.

\subsubsection{Log-Mel Spectrogram Features}

We use a standard log-Mel spectrogram as the front-end representation.
For each pre-processed waveform, STFT is first computed with a Hann window, after which the power spectrum is projected onto the Mel scale and transformed by logarithmic compression.
In the implementation, TFPARN uses \(n_{\text{fft}}=1024\), a hop length of 160 samples, and \(F=n_{\text{mels}}=160\) Mel filters, yielding a log-Mel feature tensor of shape \([T' \times F]\), where \(T'\) denotes the number of time frames.

\subsection{Test-Time Augmentation (TTA)}
\label{sec:tta}

To mitigate the variance induced by cropping randomness, we employ TTA on the input side for both the development and evaluation sets.

For utterances with duration no shorter than 4~s, the waveform is first uniformly resampled and normalized.
We then take \(K\) fixed-length 4~s windows at \(K\) evenly spaced start points, from the beginning of the utterance to its last valid start position, so that together the crops cover the whole utterance.
In our system, the TTA crop count is set to \(K=5\).

For utterances shorter than 4~s, the audio is repeated and concatenated along the time axis, after which five 4~s sub-windows are cropped from the concatenated sequence using different starting points.
Each utterance is thus represented as a multi-view waveform tensor of shape \([K, C, T]\), where \(K=5\), \(C=1\), and \(T\) denotes the number of samples in a 4~s segment.

During inference, the input tensor \([B, K, C, T]\) is reshaped into \([B\times K, C, T]\) and fed into the Transformer encoder.
Binary-classification logits are computed for each sub-window and then reshaped back to \([B, K, 2]\).

Finally, we average logits over the \(K\) views to obtain the final utterance-level output.

\section{Proposed TFPARN Model}
\label{sec:proposed_tfparn}

\subsection{Back-End Mapping Overview}

The proposed TFPARN back-end implements the following mapping chain:
\begin{itemize}
  \item frame-level embeddings,
  \item attention pooling,
  \item utterance-level embedding,
  \item binary logits.
\end{itemize}

Given an utterance \(x\) processed by the front-end, TFPARN first obtains its log-Mel spectrogram sequence
\begin{equation}
L = [l_1, l_2, \dots, l_{T'}] \in \mathbb{R}^{T' \times F},
\label{eq:logmel_seq}
\end{equation}
where \(l_t \in \mathbb{R}^{F}\) denotes the Mel spectral vector at frame \(t\), and \(T'\) is the number of frames.

The Transformer encoder then maps each frame to a hidden representation of dimension \(D\):
\begin{equation}
H = [h_1, h_2, \dots, h_{T'}] = \text{Transformer}(L) \in \mathbb{R}^{T' \times D},
\label{eq:transformer_map}
\end{equation}
where \(D\) is the hidden size.
The encoder consists of stacked multi-head self-attention layers and position-wise feed-forward networks, combined with sinusoidal positional encodings to capture long-range temporal dependencies and spoofing-related artifacts.

Given the frame-level representations \(H\), attention pooling compresses the variable-length sequence into a fixed-dimensional utterance-level embedding \(z \in \mathbb{R}^{D}\).
A lightweight scoring network assigns a weight to each frame, and \(z\) is the weighted sum of the frame representations, which lets the model emphasize frames carrying discriminative spoofing cues.
The exact formulation, together with the mean-pooling ablation variant, is given in Section~\ref{subsec:pooling}.

The utterance-level embedding \(z\) is then passed through a lightweight fully connected classification head to produce two-dimensional logits:
\begin{equation}
o = W z + b \in \mathbb{R}^{2},
\label{eq:logits}
\end{equation}
where \(o_0\) corresponds to the spoof (AI-generated) class and \(o_1\) corresponds to the bonafide (genuine human) class. 
The detection score can be derived from the logit difference or from posterior probabilities obtained via a softmax function.

During training, the classification objective adopts the focal loss formulation and is further combined with a pairwise ranking loss \cite{refFocal,refPairwise}. 
The focal loss down-weights well-classified examples and focuses training on harder samples.
The pairwise loss constructs (spoof, bonafide) pairs and enforces that bonafide scores exceed spoof scores by a predefined margin. 

\subsection{Transformer-based Encoder}

To improve numerical stability across frequency bands, we apply LayerNorm along the frequency dimension and then map each frame to the Transformer feature space of dimension \(d_{\mathrm{model}}\) via a linear projection:
\begin{equation}
\mathbf{H}_0=\mathrm{Linear}(\mathrm{LayerNorm}(\mathbf{X}_{\mathrm{mel}}))\in\mathbb{R}^{B\times T'\times d_{\mathrm{model}}}.
\end{equation}
where \(F=n_{\mathrm{mels}}=160\) and \(T'\) is the number of time frames determined by the STFT hop size (approximately \(T/\text{hop\_length}\)).
In this system, \(d_{\mathrm{model}}=256\).
We further add the standard sinusoidal positional encoding to inject frame-order information, yielding
\begin{equation}
\tilde{\mathbf{H}}_0=\mathrm{PosEnc}(\mathbf{H}_0).
\end{equation}
Based on \(\tilde{\mathbf{H}}_0\), TFPARN stacks \(L\) Transformer encoder layers, each consisting of multi-head self-attention and a feed-forward network (FFN) with residual connections and layer normalization \cite{refTransformer}:
\begin{equation}
\mathbf{H}_l=\mathrm{TransformerEncoderLayer}(\mathbf{H}_{l-1}),\quad l=1,\dots,L.
\end{equation}
We set \(L=6\), the number of attention heads to \(h=8\), the model dimension to \(d_{\mathrm{model}}=256\), and the FFN hidden dimension to \(d_{\mathrm{ff}}=1024\). 
Dropout is applied in both the attention and FFN blocks to mitigate overfitting.

\subsection{Pooling Methods}
\label{subsec:pooling}

The Transformer encoder produces a frame-level representation sequence
\begin{equation}
\mathbf{H}=\mathbf{H}_L\in\mathbb{R}^{B\times T'\times d_{\mathrm{model}}},
\end{equation}
which must be aggregated into an utterance-level embedding for binary anti-spoofing classification. 
In TFPARN, these frame-level representations are aggregated by attention pooling. 
To verify the contribution of this design, we additionally construct a mean-pooling variant for the ablation study.

\begin{itemize}
  \item \textit{Attention pooling} (default): a lightweight scoring network assigns an attention weight to each frame \cite{refAttnPool}. For each encoded vector \(\mathbf{H}_t\), we use a two-layer perceptron
  \begin{equation}
  e_t = \mathbf{w}_2^{\top}\tanh(\mathbf{W}_1\mathbf{H}_t+\mathbf{b}_1)+b_2,\quad t=1,\dots,T',
  \end{equation}
  where \(\mathbf{W}_1\in\mathbb{R}^{\frac{d_{\mathrm{model}}}{2}\times d_{\mathrm{model}}}\) and \(\mathbf{w}_2\in\mathbb{R}^{\frac{d_{\mathrm{model}}}{2}}\). The attention weights are computed by softmax normalization,
  \begin{equation}
  \alpha_t = \frac{\exp(e_t)}{\sum_{j=1}^{T'}\exp(e_j)},
  \end{equation}
  and the utterance-level representation is the weighted sum
  \begin{equation}
  \mathbf{h}_{\mathrm{utt}}=\sum_{t=1}^{T'}\alpha_t\mathbf{H}_t.
  \end{equation}
  In implementation, padding positions are assigned \(-\infty\) before softmax so that normalization is performed only over valid frames.
  
  Attention pooling emphasizes the frames most likely to carry spoofing cues while retaining global context, improving sensitivity to subtle artifacts relative to uniform averaging.

  \item \textit{Mean pooling} (ablation): we replace the attention module in the ablation experiments with a masked average over all valid frames to assess the contribution of attention pooling,
  \begin{equation}
  \mathbf{h}_{\mathrm{utt}}=\frac{1}{T'}\sum_{t=1}^{T'}\mathbf{H}_{t}.
  \end{equation}
  This approach is simple but treats all frames equally, potentially diluting sparse spoofing cues.
\end{itemize}

\subsection{Implementation Summary and Tensor Shapes}

We summarize the TFPARN backbone architecture, representative tensor shapes, and trainable parameter statistics in Tables~\ref{tab:backbone} and~\ref{tab:parameter_statistics}. 

The default attention-pooling TFPARN contains 4,846,275 (approximately 4.85M) trainable parameters. 
For the ablation study, the mean-pooling variant removes the lightweight attention scorer, reducing the total parameter count to 4,813,250 (approximately 4.81M).

\begin{table}[htbp]
\caption{TFPARN backbone architecture and tensor shapes (single utterance).}
\begin{center}
\footnotesize
\setlength{\tabcolsep}{4pt}
\renewcommand{\arraystretch}{1.05}
\resizebox{\columnwidth}{!}{%
\begin{tabular}{lcl}
\toprule
\textbf{Module / Layer} & \textbf{Output size} & \textbf{Specification} \\
\midrule
Encoder input & \(401\times 256\) & feature sequence after front-end projection \\
Transformer Encoder \(\times 6\) & \(401\times 256\) & heads=8, FFN dim=1024 \\
Attentive pooling (default) & \(256\) & \(\mathrm{Linear}(256\rightarrow128)\rightarrow\tanh\rightarrow\mathrm{Linear}(128\rightarrow1)\) \\
Mean pooling (ablation) & \(256\) & masked mean over 401 frames \\
FC1 & \(128\) & \(\mathrm{Linear}(256\rightarrow128)\) \\
ReLU + Dropout & \(128\) & dropout \(p=0.3\) \\
FC2 (logits) & \(2\) & \(\mathrm{Linear}(128\rightarrow2)\) \\
\bottomrule
\end{tabular}
}
\label{tab:backbone}
\end{center}
\end{table}

\begin{table}[htbp]
\caption{Compact trainable-parameter summary of TFPARN}
\begin{center}
\footnotesize
\setlength{\tabcolsep}{4pt}
\renewcommand{\arraystretch}{1.03}
\resizebox{\columnwidth}{!}{%
\begin{tabular}{llr}
\toprule
\textbf{Module} & \textbf{Configuration / operation} & \textbf{Trainable parameters} \\
\midrule
Input processing & LayerNorm(160) & 320 \\
Feature projection & Linear (160 $\rightarrow$ 256) & 41,216 \\
Transformer encoder $\times 6$ & 6 layers; MHSA + FFN + 2$\times$LayerNorm & 4,738,560 \\
Attention pooling & 2-layer scorer: 256 $\rightarrow$ 128 $\rightarrow$ 1 & 33,025 \\
Mean pooling (ablation) & no trainable parameters & 0 \\
Classifier & Linear (256 $\rightarrow$ 128) + Linear (128 $\rightarrow$ 2) & 33,154 \\
\midrule
\textbf{Total (attention pooling)} & --- & \textbf{4,846,275} \\
\textbf{Total (mean pooling)} & --- & \textbf{4,813,250} \\
\bottomrule
\end{tabular}
}
\label{tab:parameter_statistics}
\end{center}
\end{table}

\subsection{Focal Classification Loss}

In ASVspoof~5 Track~1, many samples become easy as training progresses, while a smaller set of confusing trials stays near the decision boundary.
Cross-entropy tends to under-emphasize these hard cases once most samples are already well classified, so TFPARN adopts focal loss to keep optimization focused on the difficult samples \cite{refFocal}.

Given the model output logits \(\mathbf{z}\in\mathbb{R}^{2}\) (class indices \(0=\mathrm{spoof}\), \(1=\mathrm{bonafide}\)), the softmax probabilities are \(\mathbf{p}=\mathrm{softmax}(\mathbf{z})\).
For a sample with ground-truth label \(y\in\{0,1\}\), we define \(p_t=p_y\).
The focal loss is formulated as
\begin{equation}
\mathcal{L}_{\mathrm{focal}} = - \alpha_y (1 - p_t)^\gamma \log p_t,
\end{equation}
where \(\alpha_y\) is the class-weight factor and \(\gamma>0\) is the focusing parameter that down-weights easy examples.

With the label encoding \(0=\mathrm{spoof}\), \(1=\mathrm{bonafide}\), the class-weight vector is \(\boldsymbol{\alpha}=[\alpha_{\mathrm{spoof}},\alpha_{\mathrm{bonafide}}]=[1-\alpha,\ \alpha]\).
We set \(\alpha=0.5\), so both classes are weighted equally; here focal loss is only used to emphasize hard samples rather than to counter class imbalance.

The focusing parameter \(\gamma\) determines how strongly the model emphasizes hard samples: when \(\gamma>0\), correctly classified samples with high confidence (large \(p_t\)) are down-weighted by the factor \((1-p_t)^\gamma\), whereas samples near the decision boundary or misclassified ones receive larger gradient signals.
We set \(\gamma=2.0\) by default.

\subsection{Pairwise Ranking Loss}

We use a pairwise ranking loss to better match EER and minDCF, which are ranking/threshold-sensitive metrics in ASVspoof~5 Track~1 \cite{refASV5}. 
These metrics assess whether bonafide scores are globally higher than spoof scores, rather than classification accuracy alone. 
We add a pairwise ranking loss alongside focal loss, which encourages a larger score margin between bonafide and spoof trials \cite{refPairwise}.

We treat the model logit for the bonafide class as the detection score \(s(\mathbf{x}) = z_1(\mathbf{x})\), where \(z_1\) denotes the logit corresponding to the bonafide class.
For each mini-batch, samples are partitioned into a bonafide set \(\mathcal{B}\) and a spoof set \(\mathcal{S}\) based on their labels. 
We construct all pairwise combinations \((b,s)\in\mathcal{B}\times\mathcal{S}\) and define a margin-based hinge ranking loss for each pair:
\begin{equation}
\ell_{\mathrm{pair}}(b,s) = \max\big(0,\, m - (s(b) - s(s))\big),
\end{equation}
where \(m>0\) is a margin hyper-parameter (we set \(m=1.0\) in this work).
The overall pairwise ranking loss is the mean over all valid pairs:
\begin{equation}
\mathcal{L}_{\mathrm{pair}} = \frac{1}{|\mathcal{B}||\mathcal{S}|} \sum_{b\in\mathcal{B}}\sum_{s\in\mathcal{S}} \ell_{\mathrm{pair}}(b,s).
\end{equation}
This loss enforces that bonafide samples obtain higher scores than spoof samples by at least the margin \(m\), which is more consistent with ROC/DET-based metrics (e.g., EER and minDCF).

\subsection{Combined TFPARN Loss Function}

Combining the above components, the overall loss computation of TFPARN consists of the main focal loss and an auxiliary pairwise ranking loss \cite{refFocal, refPairwise}. 
Let \(\mathcal{L}_{\text{focal}}\) denote the classification loss and \(\mathcal{L}_{\mathrm{pair}}\) denote the margin-based ranking loss. The total loss is
\begin{equation}
\mathcal{L}_{\text{TFPARN}} = \mathcal{L}_{\text{focal}} + \lambda \,\mathcal{L}_{\mathrm{pair}},
\end{equation}
where \(\lambda>0\) controls the relative weight of the ranking term. 
The margin hyper-parameter \(m\) in \(\mathcal{L}_{\mathrm{pair}}\) specifies the desired minimum score separation between bonafide and spoof; in this work, \(m=1.0\) and \(\lambda=0.3\) are fixed by default.

\section{Training, Scoring, and Comparison Protocol}

\subsection{Training Setup}

\subsubsection{Input Data and Labels}

This work performs end-to-end training on the ASVspoof~5 Track~1 training set \cite{refASV5}. 
The input consists of mono waveform segments resampled to 16~kHz and normalized to a fixed duration of 4.0~s. 
The labels follow a binary classification setting (\(\text{spoof} = 0, \text{bonafide} = 1\)).

The training batch size is set to 64.
In the comparative experiments, the same batch size is used for all models so that GPU memory usage is compared under a unified mini-batch setting.
Data loading uses 32 worker processes, with pinned memory and prefetch enabled.

\subsubsection{Optimizer and Learning Rate Schedule}

TFPARN is trained using the AdamW optimizer, with an initial learning rate of \(1\times10^{-4}\) and a weight decay coefficient of \(1\times10^{-2}\), to suppress overfitting and stabilize training of the Transformer backbone \cite{refAdamW}. 
For fair comparison across all systems, when a model's reference recipe uses a different batch size \(B_{\mathrm{ref}}\) and initial learning rate \(\eta_{\mathrm{ref}}\), the learning rate is adjusted proportionally according to the linear scaling rule~\cite{refLinScale} \(\eta = \eta_{\mathrm{ref}} \times 64 / B_{\mathrm{ref}}\).

The first 5 epochs use linear warmup, gradually increasing the learning rate from a small value to the base learning rate, followed by cosine annealing to gradually decay it to \(1\times 10^{-6}\) \cite{refSGDR}. 
This schedule avoids large gradient oscillations early in training and allows finer learning-rate adjustments later on.

\subsubsection{Number of Epochs and Early Stopping Strategy}

The maximum number of training epochs is set to 100. 
After each epoch, a full validation is performed on the Dev set, evaluating metrics. 
Dev minDCF is used as the early stopping criterion: when Dev minDCF does not show a significant decrease for 15 consecutive epochs (with a tolerance of approximately \(10^{-4}\)), early stopping is triggered. 

\subsubsection{Positive/Negative Sampling and Data Augmentation}

No additional class resampling or re-weighting is applied during training. 
Random shuffling is enabled only for the training set to ensure that spoof/bonafide combinations in each mini-batch are as diverse as possible.

At the waveform level, RawBoost data augmentation is applied to the training set with a probability of 0.5 \cite{refRawBoost}. 
Random cropping is applied for utterances longer than 4~s. Repeat-concatenation followed by cropping is applied for utterances shorter than 4~s. 
No online data augmentation is applied to the development or evaluation sets; only deterministic center cropping is used.

\subsection{Test-time Augmentation}

At inference, we apply TTA on the Dev and Eval sets with \(K=5\) fixed-length crops per utterance and average the per-crop logits to form the utterance-level score.
The crop-generation and fusion procedure is detailed in Subsection~\ref{sec:tta}.

\subsection{Feature Extraction and Model Input}

To ensure a fair comparison across different back-end architectures, we unified the waveform-level preprocessing and data handling protocol for all compared systems.

All models were trained and evaluated under the same waveform segmentation and sampling-rate settings.
Speech signals are resampled to 16~kHz and cropped or repeated to a fixed duration of 4.0~s.

The front-end feature extractor is model-dependent: AASIST and RawNet2 take raw waveform input, but their Sinc-like front-end filtering modules are fixed as non-trainable in our experiments; TFPARN uses log-Mel features (\(n_{\text{fft}}=1024\), \(n_{\text{mels}}=160\)).
This setting is intended to emphasize differences in the model architectures themselves and reduce the influence of front-end learning on the overall comparison.

\subsection{Evaluation Metrics}
\label{sec:eval_metrics}

We report two groups of metrics in this work: the official ASVspoof~5 Track~1 detection metrics and additional computational-efficiency metrics.
The official ASVspoof~5 metrics are minDCF, EER, Cllr, and actDCF.
Among them, minDCF is the primary challenge metric.

The following indicators are used to compare computational efficiency:
\begin{itemize}
\item Total number of trainable parameters of each model;
\item Training time per epoch;
\item Maximum GPU memory usage during inference on the Dev set;
\item Time required to reach the best Dev checkpoint;
\item Inference latency per utterance on the Dev set with TTA enabled.
\end{itemize}

All efficiency measurements are obtained under the same hardware and implementation environment.

To account for the variability introduced by random initialization and stochastic training, every system is independently trained and evaluated under three random seeds (42, 63, and 2026).
The detection metrics in Table~\ref{tab:main_results} and the efficiency metrics in Table~\ref{tab:efficiency_results} are reported as mean $\pm$ standard deviation over these three seeds (parameter counts are exact and seed-independent).

We also visualize how each metric evolves with training cost (Fig.~\ref{fig:training_efficiency}).
For each system and split we plot a detection metric against the cumulative wall-clock training time in minutes, on a log-scaled horizontal axis.
The cumulative time is accumulated only over the per-epoch training and validation passes; the one-off final evaluation pass is excluded so that its large cost does not distort the axis.
Because the systems differ by roughly an order of magnitude in total training time, we use a logarithmic axis to make the fast and slow systems share one figure without the fast ones being compressed against the left edge.
The three seeds run for different amounts of time per epoch, so their curves cannot be averaged directly. For each system, we therefore choose $30$ log-spaced time points within the range shared by all three seeds and linearly interpolate each seed's metric onto these points, so that no value is extrapolated. The solid line is the mean over the three seeds, and the shaded band is $\pm$one standard deviation.
Each curve is drawn all the way to its early-stopping point.
On the development-set panels, a star marks each seed's best (lowest) value of the metric, taken directly from the raw per-epoch records, not from the interpolated mean curve.

\section{Experiments on ASVspoof~5 Track~1 Closed Condition}

\subsection{Baseline Systems}

\begin{itemize}
  \item AASIST is one of the most widely used anti-spoofing baselines in the ASVspoof series \cite{refASV2015, refASV2017, refASV2019, refASV2021, refASV5}.
        It applies a Sinc-based convolutional front-end and a residual encoder to the raw waveform and models the encoded features with spectro-temporal graph attention. An attention-based aggregation layer then obtains a global representation, followed by a binary classification head that outputs spoof/bonafide predictions.

        In our implementation, the original Sinc-like front-end filterbank of AASIST is retained but frozen as non-trainable (no gradient updates) to reduce front-end-induced variance; the downstream convolutional and graph-based modules follow Jung et al. \cite{refAASIST}.

  \item RawNet2 is a deep model that operates directly on the raw time-domain waveform.
        Compared with feature-based systems, it often exhibits stronger robustness under certain unseen attack conditions. 
        It consists of multiple residual convolutional blocks and a global pooling layer\cite{refRawNet2}.
\end{itemize}

For fair comparison in this study, the original SincConv front-end is retained in structure but fixed as non-trainable.

For both baselines, the front-end filterbank is frozen and excluded from training, so only the back-end modules are optimized.
Table~\ref{tab:baseline_param_stats} summarizes their main components and trainable-parameter counts; the per-model totals match the parameter column of Table~\ref{tab:efficiency_results}.

\begin{table}[htbp]
\caption{Structure and trainable-parameter summary of the re-implemented baselines (front-end frozen)}
\begin{center}
\footnotesize
\setlength{\tabcolsep}{4pt}
\renewcommand{\arraystretch}{1.15}
\begin{tabular}{l>{\raggedright\arraybackslash}p{5.1cm}r}
\toprule
\textbf{System} & \textbf{Main back-end components} & \textbf{Params} \\
\midrule
AASIST & 6 residual blocks; spectro-temporal GAT and heterogeneous GAT; graph pooling; Linear$(160\rightarrow2)$ & 0.30M \\
RawNet2 & 6 residual blocks with channel attention; 3-layer GRU$(128\rightarrow1024)$; FC$(1024\rightarrow1024\rightarrow2)$ & 17.62M \\
\bottomrule
\end{tabular}
\end{center}
\label{tab:baseline_param_stats}
\end{table}

\subsubsection{Internal Experiments and Ablation Design}

Building on the two re-implemented baselines above, we construct multiple TFPARN variants to systematically examine key design choices and conduct ablation studies.
In particular, we focus on the following aspects:
\begin{itemize}
  \item \textit{Effect of the pairwise ranking branch}: under settings using only cross-entropy or focal loss, we compare the performance with and without the pairwise ranking branch, and examine its impact on EER and minDCF from the perspective of ranking consistency.
  \item \textit{Effect of the focal loss branch}: with pairwise ranking fixed on, we compare TFPARN under CE and focal loss to isolate the contribution of the focal term.
  \item \textit{Effect of attentive temporal pooling}: with fixed loss-function settings and focal loss hyperparameters \((\alpha,\gamma)\), we use the attention-pooling TFPARN as the main system and compare it with a mean-pooling ablation variant to verify the contribution of the attention-pooling design.
\end{itemize}

The detailed configurations of all systems (including the loss type, whether pairwise ranking is used, focal loss \(\alpha\) and \(\gamma\), whether RawBoost is enabled, and the pooling method) are summarized in Table~\ref{tab:system_config}. 
All results are indexed by the model IDs in this table:

\begin{table}[htbp]
\caption{System configurations for internal comparisons and ablation studies}
\label{tab:system_config}
\begin{center}
\renewcommand{\arraystretch}{1.15}
\scriptsize
\setlength{\tabcolsep}{2.6pt}
\resizebox{\columnwidth}{!}{%
\begin{tabular}{ccccc}
\toprule
\textbf{ID} & \textbf{System} & \textbf{Loss} & \textbf{Pairwise Ranking} & \textbf{Pooling} \\
\midrule
1 & AASIST  & CE         & No  & Graph Pooling \\
2 & RawNet2 & CE         & No  & Global Max \\
3 & TFPARN  & CE         & No  & Mean \\
4 & TFPARN  & CE         & Yes & Mean \\
5 & TFPARN  & \begin{tabular}{@{}l@{}}Focal Loss \(\alpha=0.5, \gamma=2.0\)\end{tabular} & Yes & Mean \\
6 & TFPARN  & \begin{tabular}{@{}l@{}}Focal Loss \(\alpha=0.5, \gamma=2.0\)\end{tabular} & Yes & Attention \\
\bottomrule
\end{tabular}}
\end{center}
\end{table}

\begin{itemize}
  \item Model 1--2 are re-implemented AASIST model and RawNet2 model;
  \item Model 3 is used for comparing the effects with the baseline models;
  \item Model 3 and Model 4 are used to compare the differences brought by pairwise ranking;
  \item Model 4 and Model 5 are used to compare the differences between CE and focal loss \cite{refFocal};
  \item Model 5 and Model 6 are used to compare the mean-pooling ablation variant with the default attention-pooling variant.
\end{itemize}

\subsection{Main Results}

Table~\ref{tab:main_results} reports the detection metrics for all six systems.

\begin{table*}[t!]
\caption{Main results of all systems (mean $\pm$ std over three seeds)}
\begin{center}
{\footnotesize
\setlength{\tabcolsep}{6pt}
\renewcommand{\arraystretch}{1.15}
\begin{tabular}{ccccc}
\toprule
\textbf{ID} & \textbf{EER(\%)} & \textbf{minDCF} & \textbf{Cllr} & \textbf{actDCF} \\
\midrule
1 & 18.58 $\pm$ 0.16 & 0.2911 $\pm$ 0.0026 & 2.6545 $\pm$ 0.8416 & 0.4966 $\pm$ 0.1648 \\
2 & 27.23 $\pm$ 0.50 & 0.5375 $\pm$ 0.0052 & 2.8672 $\pm$ 0.2752 & 0.7214 $\pm$ 0.0742 \\
3 & 12.91 $\pm$ 0.09 & 0.2662 $\pm$ 0.0045 & 1.8796 $\pm$ 0.3314 & 0.3547 $\pm$ 0.0224 \\
4 & 12.92 $\pm$ 0.07 & 0.2561 $\pm$ 0.0021 & 1.6786 $\pm$ 0.3918 & 0.3160 $\pm$ 0.0526 \\
5 & 12.70 $\pm$ 0.36 & 0.2499 $\pm$ 0.0031 & 0.7232 $\pm$ 0.0787 & 0.3325 $\pm$ 0.0211 \\
6 & 12.52 $\pm$ 0.11 & 0.2430 $\pm$ 0.0043 & 0.9243 $\pm$ 0.4907 & 0.2897 $\pm$ 0.0191 \\
\bottomrule
\end{tabular}
}
\label{tab:main_results}
\end{center}
\end{table*}

The base TFPARN (ID~3), using only cross-entropy with mean pooling, already outperforms both baselines on every metric.
Its EER of 12.91\% improves on AASIST (ID~1) at 18.58\% and RawNet2 (ID~2) at 27.23\%, and its minDCF of 0.2662 improves on their 0.2911 and 0.5375.
TFPARN has a Cllr of 1.8796 and an actDCF of 0.3547, against 2.6545 and 0.4966 for AASIST and 2.8672 and 0.7214 for RawNet2. This shows that TFPARN's advantage is even larger in calibration-dependent metrics.
TFPARN is thus a stronger detector than either baseline under the ASVspoof~5 Track~1 closed condition, even before the proposed loss and pooling changes are introduced \cite{refASV5}.

Adding the pairwise ranking branch (ID~4) leaves the EER almost unchanged, from 12.91\% to 12.92\%, but lowers minDCF, Cllr (to 1.6786) and actDCF (to 0.3160); the ranking term acts on the decision cost and score ordering rather than on the equal-error point.
Replacing cross-entropy with focal loss (ID~5) yields small improvements in EER (12.70\%) and minDCF (0.2499) and a much larger one in Cllr, which falls from 1.6786 to 0.7232, the largest single calibration improvement in the table.
Since $\alpha=0.5$ weights the two classes equally, this gain reflects the loss concentrating on hard, ambiguous utterances rather than any class rebalancing, and it is most directly reflected in Cllr, which depends on the raw score scale; actDCF moves slightly in the opposite direction, from 0.3160 to 0.3325.
Attention pooling (ID~6) yields the lowest minDCF (0.2430), EER (12.52\%) and actDCF (0.2897) of any system, but Cllr rises to 0.9243 with a wider spread across seeds.
The full model (ID~6) is the best configuration on three of the four metrics, including the primary minDCF, and ID~5 gives the lowest and most stable Cllr; both remain well ahead of the baselines.

We also compare the systems on training and inference cost.
Table~\ref{tab:efficiency_results} lists, for each system, the parameter count, peak GPU memory during inference on the Dev set, training time per epoch, time to reach the best model, and per-utterance inference latency on the Dev set.

\begin{table*}[t!]
\caption{Training and inference efficiency comparison (mean $\pm$ std over three seeds; parameter counts are exact)}
\begin{center}
{\small
\setlength{\tabcolsep}{5pt}
\renewcommand{\arraystretch}{1.15}
\resizebox{\textwidth}{!}{%
\begin{tabular}{cccccc}
\toprule
\textbf{ID} & \textbf{Parameter Count} & \textbf{Memory Usage (GB)} & \textbf{Time per Training Epoch (s/epoch)} & \textbf{Time to Best Model (min)} & \textbf{Latency per Utterance (ms/utt)} \\
\midrule
1 & 0.30M & 56.7 & 1289.4 $\pm$ 0.7 & 1014.61 $\pm$ 634.32 & 10.4805 $\pm$ 0.0034 \\
2 & 17.62M & 4.9 & 94.5 $\pm$ 1.1 & 73.77 $\pm$ 27.63 & 0.7802 $\pm$ 0.0203 \\
3 & 4.81M & 1.4 & 136.3 $\pm$ 1.8 & 207.84 $\pm$ 59.17 & 0.8073 $\pm$ 0.0542 \\
4 & 4.81M & 1.4 & 134.4 $\pm$ 0.9 & 254.70 $\pm$ 52.87 & 0.7873 $\pm$ 0.0097 \\
5 & 4.81M & 1.4 & 134.9 $\pm$ 0.4 & 167.00 $\pm$ 22.66 & 0.7893 $\pm$ 0.0040 \\
6 & 4.85M & 1.4 & 135.6 $\pm$ 0.8 & 149.40 $\pm$ 39.51 & 0.7896 $\pm$ 0.0054 \\
\bottomrule
\end{tabular}%
}
}
\label{tab:efficiency_results}
\end{center}
\end{table*}

Parameter count is a poor predictor of computational cost.
AASIST (ID~1) has the fewest trainable parameters (0.30M) but is the most expensive system to train and deploy: its heterogeneous graph attention drives peak inference memory to 56.7\,GB, training time to 1289.4\,s per epoch, and latency to 10.48\,ms per utterance, roughly 40$\times$, 10$\times$ and 13$\times$ the corresponding TFPARN figures.
RawNet2 (ID~2) is the largest model, at 17.62M parameters, but is inexpensive per epoch (94.5\,s).

The TFPARN variants have an intermediate parameter range, 4.81--4.85M, in which attention pooling (ID~6) adds only the small ($\sim$33k-parameter) scoring head over the mean-pooling variants (IDs~3--5).
They require the least memory of any system, 1.4\,GB against 4.9\,GB for RawNet2 and 56.7\,GB for AASIST, because the back-end operates on a 401-frame log-Mel sequence rather than a long raw waveform.
A TFPARN epoch takes roughly 135\,s, slower than RawNet2's 94.5\,s but faster than AASIST's 1289.4\,s, and its latency of about 0.79\,ms per utterance matches RawNet2 and is roughly 13$\times$ faster than AASIST.

Among the TFPARN variants, the full model (ID~6) converges fastest, reaching its best Dev minDCF in 149.4\,min, against 1014.6\,min for AASIST.
The system that detects best (ID~6) is therefore also the fastest TFPARN variant to converge, at a fraction of AASIST's training time and memory.

The cost of AASIST comes from its graph back-end. Each graph-attention layer builds a full tensor over all pairs of nodes and then projects and normalizes it. AASIST repeats this many times: in both a spectral and a temporal branch, across four heterogeneous graph-attention layers on two parallel paths, and in several top-$k$ graph-pooling stages. All of them operate on feature maps taken from the raw 64{,}000-sample waveform.
These operations are irregular and create many intermediate tensors, so they run far less efficiently on the GPU than plain matrix multiplications. This is why a model with only 0.30M parameters is the slowest system in the comparison and the one that uses the most memory.
TFPARN, in contrast, is built almost entirely from standard operations that GPUs handle well---six Transformer encoder layers, dense linear projections, and a small pooling and classification head---applied to a short 401-frame log-Mel sequence.
Self-attention grows quadratically with sequence length, but with only 401 frames this cost stays small, and its regular dense computation is exactly what current GPU libraries run most efficiently.

TFPARN spends modestly more time per epoch (about 135\,s against 94.5\,s) but converts it into a more discriminative representation.
Both systems use a Mel-scale front-end, yet RawNet2 feeds only 20 frozen SincConv bands into a 1-D convolutional residual stack and a three-layer recurrent network, whereas TFPARN forms a 160-band log-Mel spectrogram and models it with global self-attention.
In this explicit time--frequency representation, localized spoofing cues---narrow-band irregularities, unnatural harmonic structure, and broken spectro-temporal texture---are easy to access directly, instead of having to be recovered from the raw waveform by a recurrent network.
This is consistent with the training-efficiency curves discussed next, where RawNet2 attains the lowest training error of the three systems yet generalizes worst, while TFPARN's additional per-epoch cost buys a representation that transfers to unseen data.

\begin{figure*}[tp]
  \centering
  \subfloat[minDCF, training set]{\includegraphics[width=0.40\textwidth]{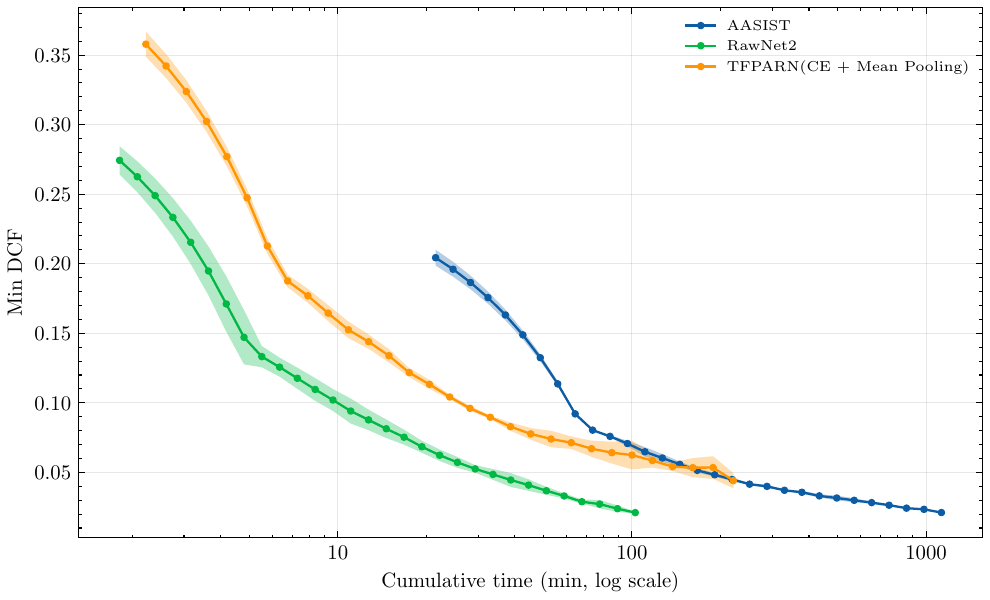}\label{fig:eff_train_mindcf}}
  \hfil
  \subfloat[minDCF, development set]{\includegraphics[width=0.40\textwidth]{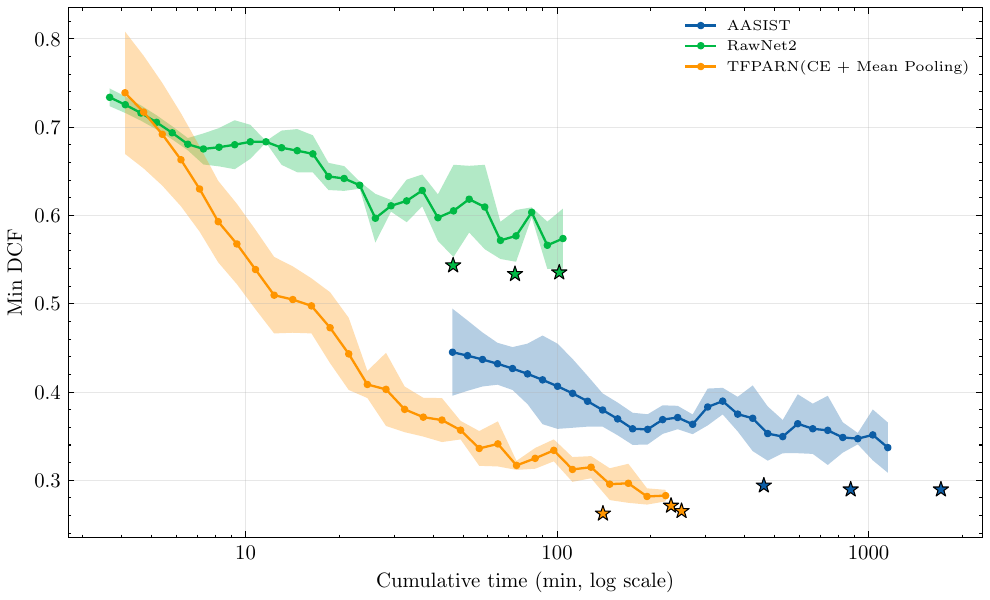}\label{fig:eff_dev_mindcf}}
  \\
  \subfloat[EER, training set]{\includegraphics[width=0.40\textwidth]{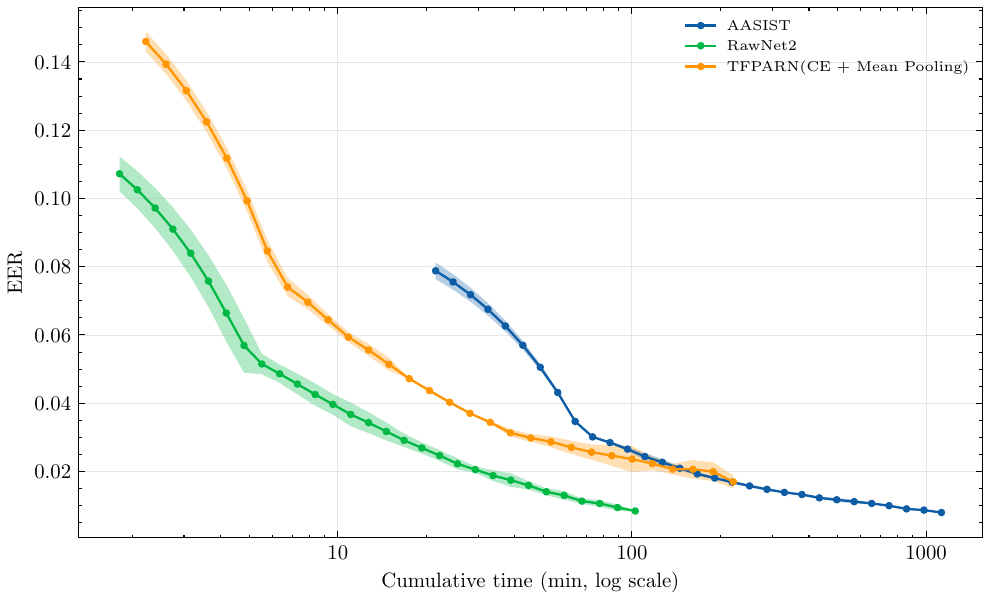}\label{fig:eff_train_eer}}
  \hfil
  \subfloat[EER, development set]{\includegraphics[width=0.40\textwidth]{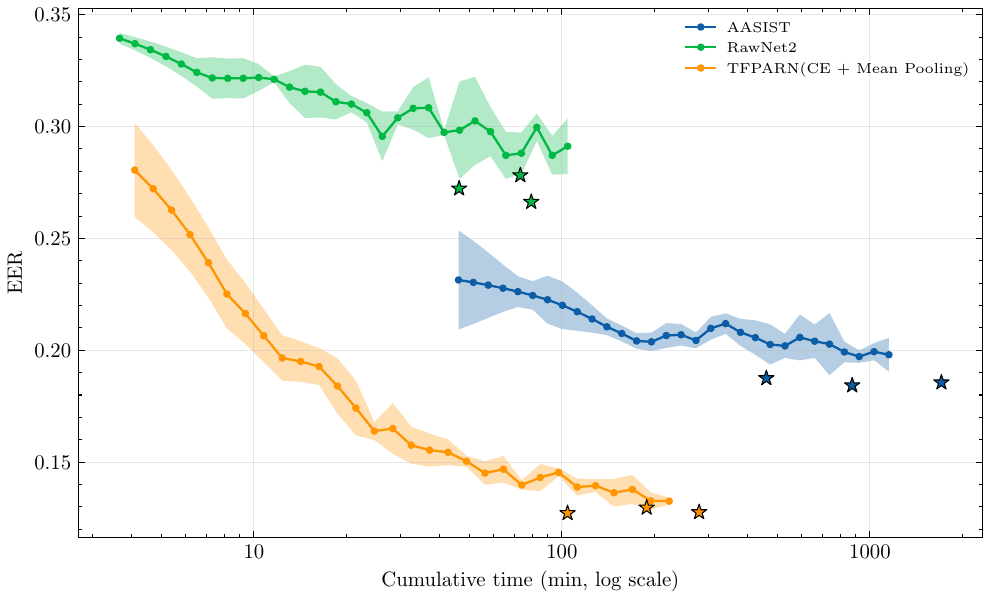}\label{fig:eff_dev_eer}}
  \\
  \subfloat[Cllr, training set]{\includegraphics[width=0.40\textwidth]{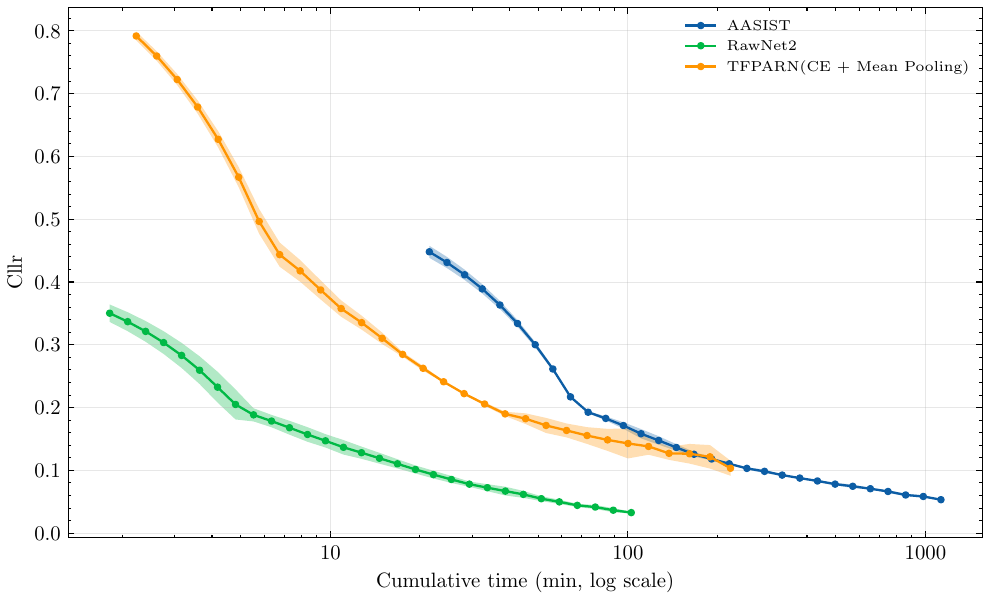}\label{fig:cal_train_cllr}}
  \hfil
  \subfloat[Cllr, development set]{\includegraphics[width=0.40\textwidth]{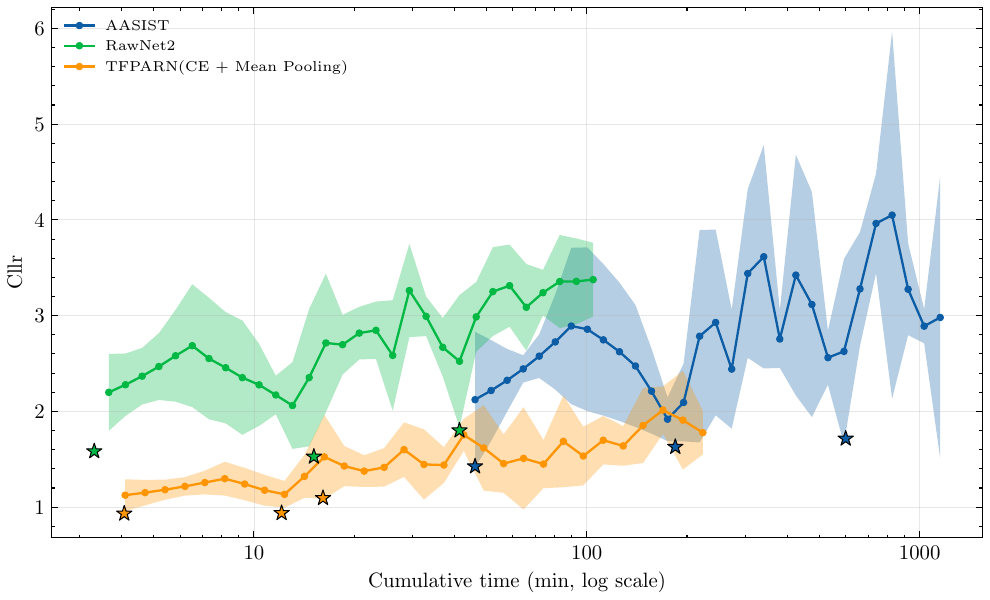}\label{fig:cal_dev_cllr}}
  \\
  \subfloat[actDCF, training set]{\includegraphics[width=0.40\textwidth]{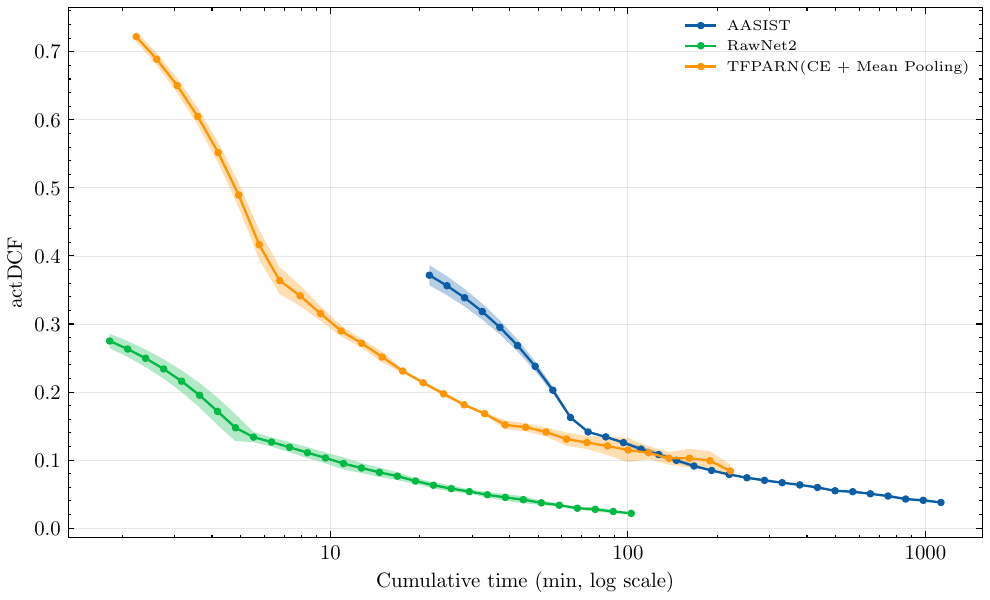}\label{fig:cal_train_actdcf}}
  \hfil
  \subfloat[actDCF, development set]{\includegraphics[width=0.40\textwidth]{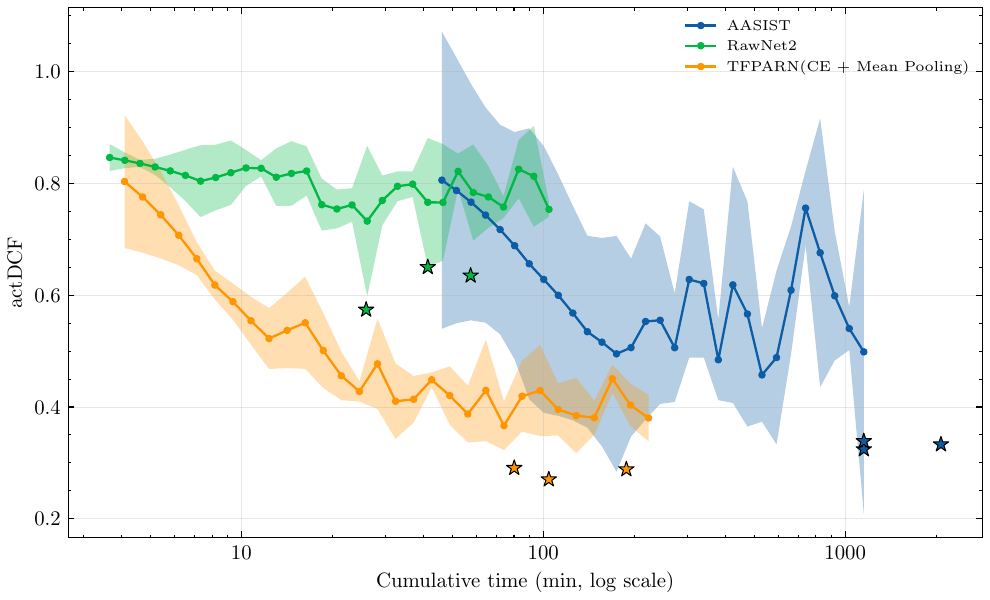}\label{fig:cal_dev_actdcf}}
  \caption{Training efficiency and generalization of the three core systems---ID~1 (AASIST), ID~2 (RawNet2), and ID~3 (TFPARN with cross-entropy loss and mean pooling)---each metric shown versus cumulative wall-clock training time (log scale). Rows, top to bottom: minDCF, EER, Cllr, actDCF (the upper two are ranking/threshold metrics, the lower two are calibration-dependent). Left column: training set; right column: development set. Solid lines are the across-seed mean (seeds 42, 63, 2026), shaded bands $\pm$one standard deviation; stars on the development-set panels mark each seed's lowest value. Lower is better.}
  \label{fig:training_efficiency}
\end{figure*}

Figure~\ref{fig:training_efficiency} plots all four metrics against cumulative wall-clock training time for the three core systems: the baselines ID~1 (AASIST) and ID~2 (RawNet2), and the base TFPARN (ID~3, cross-entropy with mean pooling) on which the other variants build.
The top two rows are the ranking/threshold metrics (minDCF, EER) and the bottom two the calibration-dependent ones (Cllr, actDCF), with the training set on the left and the development set on the right.
The curves are constructed as described in Section~\ref{sec:eval_metrics}.

On the development set (right column), TFPARN (ID~3) has the lowest minDCF (Fig.~\ref{fig:eff_dev_mindcf}) and the lowest EER (Fig.~\ref{fig:eff_dev_eer}), and it does so after far less training time than AASIST (ID~1).
AASIST's high per-epoch cost shifts its entire curve roughly an order of magnitude to the right, and even at convergence it does not reach TFPARN's Dev minDCF.
RawNet2 (ID~2) is inexpensive per epoch but plateaus at a clearly worse minDCF and EER.
This is the central efficiency claim of this work: TFPARN achieves stronger detection performance while requiring less training time.

The training-set panels (left column) show that this advantage does not stem merely from fitting the training data more closely.
RawNet2 in fact has the lowest training minDCF and EER of the three, yet generalizes the worst on the development set, which is a clear sign of overfitting.
TFPARN maintains a much smaller gap between its training and development curves, indicating that its development-set advantage reflects better generalization.

Unlike minDCF and EER, Cllr and actDCF depend on the actual score scale and threshold, so they indicate whether a system's raw scores, not merely their ordering, transfer to unseen data.
On the development set, the behavior diverges.
Only TFPARN maintains a steady downward trend.
RawNet2's development curves improve only marginally and remain high, whereas AASIST's are highly unstable, with a very wide $\pm$std band.
TFPARN is the only system whose calibration metrics hold up on the development set.
This matches its minDCF and EER behavior and shows that its advantage extends beyond ranking to the scores.

\section{Conclusion}

We presented TFPARN, a Transformer-based countermeasure for the ASVspoof~5 Track~1 closed condition that pairs an attentive temporal pooling back-end with a combined focal--pairwise training objective.
The design aims for a balance between detection performance and computational cost, not for detection accuracy alone. Log-Mel frames are encoded by a six-layer Transformer and aggregated by a lightweight attention head. Training combines a focal classification loss, which focuses on hard trials, with a pairwise ranking term that aligns the objective with the ranking- and threshold-sensitive metrics of the challenge.

Under a unified protocol against re-implemented AASIST and RawNet2 baselines, the base TFPARN already outperforms both on every detection metric, and the ablation chain lowers the primary minDCF monotonically as each component is added: $0.2662 \rightarrow 0.2561 \rightarrow 0.2499 \rightarrow 0.2430$.
The full model attains the lowest minDCF (0.2430), EER (12.52\%) and actDCF (0.2897), while the focal-loss variant gives the lowest and most stable Cllr (0.7232); the pairwise term acts mainly on decision cost and score ordering rather than on the equal-error point, and the focal term most clearly improves the calibration-dependent metrics.
These gains come at a low computational cost.
The TFPARN variants require only 1.4\,GB of peak inference memory, against 4.9\,GB for RawNet2 and 56.7\,GB for AASIST, take roughly 135\,s per epoch, and run at about 0.79\,ms per utterance; the full model also reaches its best development checkpoint fastest among the variants.
The training-efficiency curves further indicate that this advantage reflects better generalization: RawNet2 attains the lowest training-set error yet generalizes worst, whereas TFPARN maintains a small gap between its training and development behavior on both ranking and calibration metrics.
TFPARN therefore matches or surpasses both baselines on detection while training and running at substantially lower cost.

The limitation in our work is the pairwise ranking component. It is a simple hinge loss over bonafide--spoof pairs sampled within each mini-batch in our implementation, so it does little more than push scores toward a correct relative ordering.
The resulting objective is only loosely tied to minDCF and actDCF: it approximates the desired ranking rather than directly minimizing the detection cost that these metrics measure.
Replacing the pairwise hinge with a more stable listwise or differentiable ranking loss, defined over the full batch of scores rather than isolated pairs, would track the evaluation metrics more closely and make training more directly cost-sensitive.

\end{document}